\documentclass{article}


\usepackage[utf8]{inputenc}
\usepackage[T1]{fontenc}
\usepackage{hyperref}
\usepackage{url}
\usepackage{booktabs}
\usepackage{amsfonts}
\usepackage{amsmath}
\usepackage{amssymb}
\usepackage{nicefrac}
\usepackage{microtype}
\usepackage{xcolor}
\usepackage{graphicx}
\usepackage{algorithm}
\usepackage{algorithmic}
\usepackage{tikz}
\usetikzlibrary{shapes.geometric,arrows.meta,positioning,fit,backgrounds}

\title{SIR-Bench: Evaluating Investigation Depth in\\Security Incident Response Agents}

\author{
  Daniel Begimher\thanks{Equal contribution.} \quad
  Cristian Leo\footnotemark[1] \quad
  Jack Huang\footnotemark[1] \quad
  Pat Gaw \quad
  Bonan Zheng \\[4pt]
  Amazon Web Services
}

\begin{document}

\maketitle

\begin{abstract}
We present SIR-Bench, a benchmark of 794 test cases for evaluating autonomous security incident response agents that distinguishes genuine forensic investigation from alert parroting. Derived from 129 anonymized incident patterns with expert-validated ground truth, SIR-Bench measures not only whether agents reach correct triage decisions, but whether they discover novel evidence through active investigation. To construct SIR-Bench, we develop Once Upon A Threat (OUAT), a framework that replays real incident patterns in controlled cloud environments, producing authentic telemetry with measurable investigation outcomes. Our evaluation methodology introduces three complementary metrics: triage accuracy ($M_1$), novel finding discovery ($M_2$), and tool usage appropriateness ($M_3$), assessed through an adversarial LLM-as-Judge that inverts the burden of proof---requiring concrete forensic evidence to credit investigations. Evaluating our SIR agent on the benchmark demonstrates 97.1\% true positive (TP) detection, 73.4\% false positive (FP) rejection, and 5.67 novel key findings per case, establishing a baseline against which future investigation agents can be measured.
\end{abstract}

\section{Introduction}

Security Operations Centers (SOCs) face a fundamental scalability challenge: cloud environments generate millions of telemetry events daily, yet human analysts can only thoroughly investigate a fraction of the resulting alerts. Foundation models---including Large Language Models (LLMs) and emerging reasoning models---offer a potential solution through autonomous investigation agents~\cite{ircopilot}, but rigorous evaluation is needed before deploying such systems, yet existing benchmarks remain inadequate for this task.

A core challenge in evaluating investigation agents is that investigation quality itself resists simple measurement. Human analyst performance varies across individuals and incidents, and expert evaluation is expensive and difficult to standardize. With SIR-Bench's construction methodology---real incidents replayed as controlled simulations with expert-validated key findings---we produce evaluations that strongly correlate with human expert assessment while remaining reproducible and scalable.

To achieve evaluation that aligns with human investigative behavior, our methodology mimics the analyst's working environment---introducing the same signals, evidence trails, and telemetry that human investigators rely on. Knowledge benchmarks test domain understanding, not investigative reasoning~\cite{secqa,dfirmetric}. Capture The Flag (CTF)-style evaluations reward finding predetermined flags, but real investigations have no flags---only evidence that must be discovered and correlated. Standard classification metrics primarily measure whether an agent reaches the correct triage decision, but provide limited insight into whether it performed genuine investigation to justify that conclusion. By grounding evaluation in controlled attack simulations with expert-validated findings, SIR-Bench measures whether agents perform genuine forensic investigation, not merely pattern recognition.

This distinction matters because security investigations are uniquely susceptible to \textit{alert parroting}: an agent that restates alert content without discovering new evidence. A genuine investigation discovers novel findings: the exfiltrated credentials assumed cross-account roles, accessed specific S3 buckets, transferred data to external IPs.

Evaluating investigation quality requires two capabilities that existing benchmarks lack. First, evaluation must be grounded in realistic telemetry from actual attack executions---without authentic evidence trails, benchmarks cannot distinguish genuine investigation from surface-level pattern matching. We address this through OUAT, which simulates real incident patterns in controlled cloud environments, producing authentic CloudTrail artifacts with expert-validated ground truth. Second, agents may reach correct triage decisions through flawed reasoning, or demonstrate sound methodology while missing key findings. Directly evaluating reasoning is prohibitively expensive at scale. Instead, we summarize investigation outcomes as discrete key findings and evaluate whether agents surface these findings---capturing investigation quality through measurable outputs rather than costly reasoning analysis.

We address these through SIR-Bench, a benchmark of 794 test cases derived from 129 anonymized incident patterns, and an adversarial evaluation methodology that defaults to ``no security activity'' unless investigations provide concrete evidence beyond alert details.

\section{Prior Work}

\textbf{Security AI Benchmarks.} Existing benchmarks evaluate security \textit{knowledge} or \textit{offensive} capabilities, but not defensive investigation. SecQA~\cite{secqa} and DFIR-Metric~\cite{dfirmetric} assess forensics knowledge through Q\&A and CTF tasks---testing what LLMs \textit{know}, not whether they can \textit{conduct} investigations. CAIBench~\cite{caibench} and PACEbench~\cite{pacebench} evaluate offensive exploitation capabilities. CyberTeam~\cite{cyberteam} benchmarks \textit{proactive} threat hunting, whereas we evaluate \textit{reactive} incident investigation with specific evidence trails. None assess end-to-end investigation on real incidents requiring triage decisions and novel finding discovery.

\textbf{LLM-as-Judge.} The LLM-as-Judge paradigm~\cite{llmasjudge} enables scalable evaluation but raises reliability concerns. Haldar and Hockenmaier~\cite{ratingroulette} demonstrate low intra-rater reliability---judges produce variable scores on identical inputs. For security evaluation, confirmation bias manifests as judges accepting alert repetition as valid investigation~\cite{llmasjudgese}. We address this through adversarial judge design: assuming no security activity occurred and requiring concrete evidence to change this assessment, inverting the burden of proof.

\textbf{Automated IR Systems.} Security Orchestration, Automation and Response (SOAR) platforms~\cite{soar} automate playbook execution but lack adaptive reasoning. IRCopilot~\cite{ircopilot} proposes multi-LLM architectures for incident response. Our contribution is complementary---we provide the evaluation framework to rigorously assess such systems.

\section{Methodology}

To evaluate the performance of our SIR agent against true positive incidents, we prepared realistic customer cloud environments with simulated security incidents to investigate. This required first deploying benign cloud architectures that replicate customer deployments followed by executing sequences of malicious actions against the environment based on real observed attack patterns. The following section describes our dataset distribution and design as well as our methodology for data generation.

\subsection{Dataset Distribution and Design}

To ground our attack simulations, we derived 129 incident patterns from anonymized and sanitized internal security response records, removing all customer-identifiable information. These patterns---representing attack sequences, environmental configurations, and telemetry characteristics---were then used to seed 794 controlled simulations with synthetic variations, ensuring no original customer data appears in the benchmark. When the source investigation was a false positive (i.e., benign activity that triggered an alert), we converted it into a true positive by replaying actual attack actions within the same environmental context, ensuring the telemetry contained genuine indicators of compromise.

We categorized investigations across four common patterns based on attack type: Brute Force, Malicious File Execution, Misconfiguration, and Unauthorized Access. Unlike CTF-based benchmarks~\cite{dfirmetric,caibench} that use synthetic scenarios with clean attack signals, our categories reflect real-world incident distributions, where investigations involve noisy telemetry, overlapping benign activity, and incomplete evidence---conditions that more accurately measure an agent's operational readiness. Table~\ref{tab:dataset} summarizes the final dataset composition.

\begin{table}[h]
\centering
\caption{SIR-Bench dataset composition}
\label{tab:dataset}
\begin{tabular}{@{}lrrrrr@{}}
\toprule
\textbf{Attack Category} & \textbf{TP Cases} & \textbf{FP Cases} & \textbf{Total} & \textbf{\%} \\
\midrule
Brute Force & 135 & 59 & 194 & 24.4\% \\
Unauthorized Access & 186 & 112 & 298 & 37.5\% \\
Misconfiguration & 100 & 75 & 175 & 22.0\% \\
Malicious File Execution & 54 & 73 & 127 & 16.0\% \\
\midrule
\textbf{Total} & \textbf{475} & \textbf{319} & \textbf{794} & \textbf{100\%} \\
\bottomrule
\end{tabular}
\end{table}

The category distribution reflects production incident prevalence: Unauthorized Access represents 37.5\% of the evaluation set, matching observed prevalence in historical true positive cases. The overall false positive rate (40.2\%) approximates production alert characteristics where security monitoring systems generate substantial benign alerts requiring analyst triage.

\subsection{Once Upon A Threat (OUAT): GenAI-powered Attack Simulation}

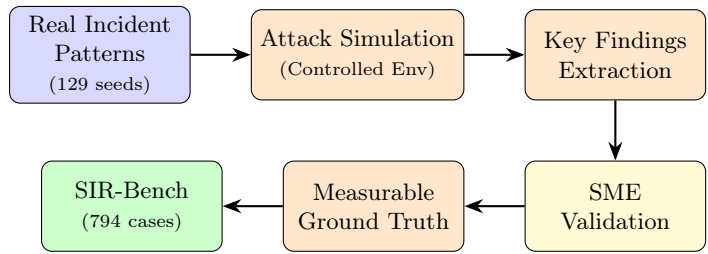
\begin{figure}[t]
\centering
\begin{tikzpicture}[
    node distance=0.5cm and 0.8cm,
    box/.style={rectangle, draw, rounded corners, minimum height=1.2cm, minimum width=2.4cm, align=center, font=\small},
    arrow/.style={-{Stealth[length=2.5mm]}, thick}
]

% Top row (left to right)
\node[box, fill=blue!15] (patterns) {Real Incident\\Patterns\\{\scriptsize(129 seeds)}};
\node[box, fill=orange!20, right=of patterns] (simulation) {Attack Simulation\\{\scriptsize(Controlled Env)}};
\node[box, fill=orange!20, right=of simulation] (extraction) {Key Findings\\Extraction};

% Bottom row (right to left)
\node[box, fill=yellow!20, below=0.8cm of extraction] (sme) {SME\\Validation};
\node[box, fill=orange!20, left=of sme] (groundtruth) {Measurable\\Ground Truth};
\node[box, fill=green!20, left=of groundtruth] (sirbench) {SIR-Bench\\{\scriptsize(794 cases)}};

% Arrows - top row
\draw[arrow] (patterns) -- (simulation);
\draw[arrow] (simulation) -- (extraction);

% Arrow - down
\draw[arrow] (extraction) -- (sme);

% Arrows - bottom row
\draw[arrow] (sme) -- (groundtruth);
\draw[arrow] (groundtruth) -- (sirbench);

\end{tikzpicture}
\caption{OUAT pipeline: Real incident patterns seed attack simulations in controlled environments. Key findings are extracted, validated by security Subject Matter Experts (SMEs), and transformed into measurable ground truth for the SIR-Bench dataset.}
\label{fig:ouat}
\end{figure}

Generating realistic security benchmarks requires authentic attack telemetry---but production incident data is sensitive and purely synthetic scenarios risk oversimplification. We developed Once Upon A Threat (OUAT) to bridge this gap: a framework that takes real incident patterns and replays them in controlled cloud environments, producing authentic CloudTrail logs and actual AWS resources for triage with expert-validated ground truth. Figure~\ref{fig:ouat} illustrates the pipeline.

OUAT begins with a seed incident from our anonymized responder dataset. The framework first provisions a realistic cloud environment matching the original architecture: EC2 instances with appropriate profiles, S3 buckets with varying access policies, IAM roles with realistic permission boundaries, and supporting services as required. Infrastructure is provisioned by executing boto3 python scripts using benign credentials, enabling consistent reproduction. Table~\ref{tab:ouat} summarizes the infrastructure and attack actions for each category.

\begin{table}[h]
\centering
\caption{OUAT attack simulation components by category}
\label{tab:ouat}
\begin{tabular}{@{}p{2.2cm}p{4.8cm}p{4.8cm}@{}}
\toprule
\textbf{Category} & \textbf{Infrastructure Provisioned} & \textbf{Attack Actions Simulated} \\
\midrule
Brute Force & IAM users, CloudTrail logging, authentication endpoints & Failed auth attempts, credential stuffing, successful compromise \\
\addlinespace
Unauthorized Access & IAM roles, S3 buckets, cross-account trust relationships & Role assumption chains, S3 enumeration, data exfiltration \\
\addlinespace
Misconfiguration & S3 buckets, security groups, IAM policies & Public exposure, overly permissive policies, unintended access \\
\addlinespace
Malicious File Exec. & EC2 instances, SSM agents, security groups & Malware download, reverse shells, persistence mechanisms \\
\bottomrule
\end{tabular}
\end{table}

With the environment ready, OUAT executes the attack sequence derived from the seed incident's timeline. For a credential compromise scenario, this might involve simulating leaked access keys, followed by cross-account role assumptions, S3 object enumeration, and data exfiltration to external endpoints. Actions execute with realistic timing---compressed from the original timeline but preserving causal ordering. Moreover, these attacks are executed from a dedicated Kali instance hosted in another AWS account to simulate true external attack conditions.

The final phase extracts ground truth. OUAT automatically correlates executed actions with generated CloudTrail events and artifacts to label expected findings. Security engineers then review and augment these labels, adding findings that require inference (e.g., recognizing ``attacker established persistence'' from observing IAM user creation followed by access key generation) and removing findings unreasonable to expect from available telemetry. Thus, by collecting key findings through both OUAT and SME knowledge, we establish a realistic ground truth that reflects what a strong, thorough investigation should entail.

To scale from 129 seed cases to 794 test cases, OUAT employs LLM-assisted variation. Given a seed case, a frontier LLM generates variations that preserve attack semantics while modifying surface details: different resource names, regional variations, temporal shifts, and alternative techniques achieving similar objectives. Each variation undergoes security engineer review; approximately 15\% are rejected due to unrealistic sequences or insufficient forensic artifacts.

False positive generation requires particular care---cases must appear suspicious enough to trigger alerts while being definitively benign upon investigation. OUAT generates these through legitimate administrative activity that matches attack patterns, intentionally public resources that trigger exposure alerts, CI/CD pipeline artifacts that resemble suspicious API patterns, and cross-account access from partner integrations. For each, investigation reveals clear evidence of benign intent.

\subsection{Ground Truth Labeling Process}

Each test case includes expert-labeled ground truth comprising:

\textbf{Triage decision} $y^* \in \{\text{TP}, \text{FP}\}$: The correct classification based on whether the simulated activity represents genuine malicious behavior or benign operations.

\textbf{Expected findings} $F^*$: The complete set of security-relevant facts an investigation should identify, including both alert-derivable information and novel discoveries.

\textbf{Novel findings} $F^*_{\text{novel}} \subseteq F^*$: The subset of findings requiring active investigation beyond alert content. These are tagged with the tool(s) required for discovery (e.g., ``CloudTrail:LookupEvents'', ``IAM:ListRolePolicies'').

\textbf{Evidence artifacts}: Specific CloudTrail event IDs, resource ARNs, and timestamps that constitute forensic evidence for each finding.

Labeling was performed by a team of five security engineers with 3--8 years of incident response experience. Each case was labeled by one engineer and reviewed by a second; disagreements (approximately 8\% of cases) were resolved through discussion. Inter-annotator agreement on triage decisions was 94.2\% ($\kappa = 0.87$); agreement on novel finding sets averaged 81.3\% Jaccard similarity.

\section{Evaluation Methodology}

A core challenge in evaluating security investigation agents is ensuring automated metrics reflect human expert judgment. We designed our evaluation framework with human alignment as the foundational principle: every automated score must correlate strongly with how security engineers assess investigation quality.

To establish this alignment, we conducted a calibration study with our security team who independently scored randomly sampled investigation reports. Each analyst marked ground truth findings as matched or unmatched based on agent output, and assessed whether triage decisions were supported by sufficient evidence. Inter-annotator agreement was 94.2\% ($\kappa = 0.87$), establishing the upper bound for automated evaluation. We then optimized our automated scoring to maximize agreement with these human annotations, achieving 91.3\% concordance at the finding level.

This human-aligned framework enables two primary metrics: $M_1$ measures triage accuracy (correct classification) and $M_2$ measures investigation depth (novel finding discovery). We include $M_3$ (tool usage appropriateness) as an optional metric---valuable for evaluating smaller models but less differentiating for frontier LLMs that achieve near-perfect tool coverage.

\subsection{$M_1$: Triage Accuracy}

The primary objective of SIR-Bench is correct triage: determining whether an alert represents genuine malicious activity requiring response (true positive) or benign behavior that can be dismissed (false positive). We formalize triage accuracy through two component metrics measuring detection and rejection rates:

\begin{equation}
M1_{\text{TP}} = \frac{|\{i : y_i = \text{TP} \land y^*_i = \text{TP}\}|}{|\{i : y^*_i = \text{TP}\}|}
\end{equation}

\begin{equation}
M1_{\text{FP}} = \frac{|\{i : y_i = \text{FP} \land y^*_i = \text{FP}\}|}{|\{i : y^*_i = \text{FP}\}|}
\end{equation}

$M1_{\text{TP}}$ measures detection rate (recall)---the fraction of actual security incidents correctly identified. $M1_{\text{FP}}$ measures false alarm rejection rate (precision proxy)---the fraction of benign alerts correctly dismissed. Neither metric alone suffices: an agent achieving 100\% $M1_{\text{TP}}$ by classifying everything as a true positive would have 0\% $M1_{\text{FP}}$, providing no operational value.

To capture the trade-off between detection and rejection in a single metric, we adopt the $F_\beta$ score:

\begin{equation}
M1_{F_\beta} = (1 + \beta^2) \cdot \frac{M1_{\text{TP}} \cdot M1_{\text{FP}}}{\beta^2 \cdot M1_{\text{TP}} + M1_{\text{FP}}}
\end{equation}

The $\beta$ parameter encodes the relative cost of missing a true positive versus raising a false alarm. We set $\beta = 3$, weighting recall nine times more heavily than precision. This choice reflects two considerations specific to LLM-based agents:

\textbf{First, the asymmetric security cost:} a missed true positive allows an attacker to maintain access, potentially leading to data exfiltration, lateral movement, or persistent compromise. A false alarm wastes analyst time but causes no direct security harm.

\textbf{Second, and more critically for LLM evaluation, the asymmetric likelihood of error modes.} When an agent incorrectly classifies a benign alert as malicious ($y = \text{TP}, y^* = \text{FP}$), this typically reflects conservative behavior---the agent observed ambiguous signals and erred toward caution. When an agent incorrectly dismisses a real attack ($y = \text{FP}, y^* = \text{TP}$), this more likely reflects hallucination or superficial reasoning---the agent failed to discover or correctly interpret evidence of malicious activity. The $\beta = 3$ weighting penalizes the latter failure mode more heavily, as it indicates a more fundamental capability gap.

We establish human baseline performance from SOC operational data. Tier-2 security analysts achieve approximately 85--90\% true positive detection with 70--80\% false positive rejection, yielding $M1_{F_3} \approx 0.86$. These baselines reflect realistic human performance under time pressure with incomplete information.

\subsubsection{$M_1$ Results}

All results are averaged over 10 independent runs conducted over a two-week evaluation period, with standard deviations ranging from 1--3 percentage points across metrics. Table~\ref{tab:m1results} presents triage accuracy results comparing the SIR agent against human analyst baselines.

\begin{table}[h]
\centering
\caption{Triage accuracy ($M_1$) results}
\label{tab:m1results}
\begin{tabular}{@{}lccccc@{}}
\toprule
& $M1_{\text{TP}}$ & $M1_{\text{FP}}$ & $M1_{F_3}$ & Missed TPs & False Alarms \\
\midrule
\textbf{SIR Agent} & \textbf{97.1\%} & \textbf{73.4\%} & \textbf{0.942} & 14/475 & 85/319 \\
\bottomrule
\end{tabular}
\end{table}

The SIR agent achieves 97.1\% true positive detection while maintaining 73.4\% false positive rejection. For context, SOC operational data suggests Tier-2 security analysts typically achieve 85--90\% true positive detection and 70--80\% false positive rejection under time pressure with incomplete information. The agent's false positive rejection falls within this human analyst range, while its 97.1\% detection rate demonstrates the potential for AI-assisted investigation to augment analyst workflows, particularly in high-volume alert environments.

Table~\ref{tab:m1category} breaks down triage performance by attack category, revealing significant variation.

\begin{table}[h]
\centering
\caption{Triage accuracy by attack category}
\label{tab:m1category}
\begin{tabular}{@{}lcccc@{}}
\toprule
\textbf{Attack Category} & $M1_{\text{TP}}$ & $M1_{\text{FP}}$ & $M1_{F_3}$ & \textbf{n} \\
\midrule
Brute Force & 99.3\% & 61.0\% & 0.958 & 194 \\
Unauthorized Access & 97.3\% & 76.8\% & 0.951 & 298 \\
Misconfiguration & 94.0\% & 82.7\% & 0.929 & 175 \\
Malicious File Execution & 96.3\% & 47.8\% & 0.891 & 127 \\
\midrule
\textbf{Overall} & \textbf{97.1\%} & \textbf{73.4\%} & \textbf{0.942} & \textbf{794} \\
\bottomrule
\end{tabular}
\end{table}

Brute Force cases achieve the highest accuracy ($M1_{F_3} = 0.958$), likely due to clear authentication failure patterns in CloudTrail. Malicious File Execution shows the lowest performance ($M1_{F_3} = 0.891$), reflecting limited telemetry for instance-level activity without additional logging sources.

\subsection{$M_2$: Investigation Depth}

Triage accuracy measures whether an agent reaches the correct conclusion, but not whether it performed genuine investigation to justify that conclusion. An agent could achieve high $M_1$ by learning statistical patterns---``credential exfiltration alerts are usually true positives''---without examining any evidence. Unlike knowledge-based benchmarks that accept correct answers regardless of reasoning~\cite{secqa}, we require agents to demonstrate genuine forensic capability. To distinguish genuine forensic investigation from superficial classification, we introduce investigation depth metrics based on finding discovery.

We partition findings into two categories aligned with our problem formalization. Let $F$ denote the set of all findings reported by the agent, and $F_{\text{novel}} \subseteq F$ denote the subset of novel findings requiring investigation beyond alert content. A finding $f \in F$ is classified as novel if it satisfies three criteria: (1) $f$ was not present in the initial alert description $d$, (2) $f$ provides independent evidence of security activity or its absence, and (3) discovering $f$ required querying tools in $\mathcal{T}$ rather than reformulating alert content.

For each test case, security engineers label ground truth findings $(F^*, F^*_{\text{novel}})$. To determine whether agent-reported findings match ground truth, we employ ROUGE-L scoring (longest common subsequence overlap) rather than LLM-based judgment. We evaluated LLM-based matching in preliminary experiments but found that human annotators frequently disagreed with the LLM's match decisions; ROUGE-L was ultimately preferred for its speed, low cost, and deterministic properties. Each ground truth finding $f^* \in F^*$ is compared against all claims in the agent's report; a match is recorded if ROUGE-L exceeds threshold $\tau = 0.42$, selected by sweeping thresholds against human-annotated match/no-match labels and choosing the value that maximized agreement (91.3\%).

We then define recall metrics for both finding types:

\begin{equation}
M2_{\text{recall}} = \frac{|F \cap F^*|}{|F^*|}
\quad\quad
M2_{\text{novel-recall}} = \frac{|F_{\text{novel}} \cap F^*_{\text{novel}}|}{|F^*_{\text{novel}}|}
\end{equation}

The distinction between regular and novel findings captures investigation quality at different levels. High $M2_{\text{recall}}$ with low $M2_{\text{novel-recall}}$ indicates an agent that identifies obvious findings (often derivable from the alert itself) but fails to conduct deeper investigation. High $M2_{\text{novel-recall}}$ indicates genuine forensic capability---the agent discovered evidence that required tool usage and reasoning beyond alert content.

We additionally report threshold metrics to characterize the distribution of investigation depth:

\begin{equation}
M2_{\text{threshold}}(N) = \frac{|\{i : |F_{\text{novel},i}| \geq N\}|}{|\text{cases}|}
\end{equation}

This measures the fraction of cases where the agent discovered at least $N$ novel findings, providing insight into consistency of deep investigation across the benchmark.

\subsubsection{$M_2$ Results}

Table~\ref{tab:m2results} presents investigation depth results comparing the SIR agent against human analyst baselines.

\begin{table}[h]
\centering
\caption{Investigation depth ($M_2$) results}
\label{tab:m2results}
\begin{tabular}{@{}lccccc@{}}
\toprule
& Avg. Novel KF & Novel Coverage & Hit 5+ KF & Hit 7+ KF \\
\midrule
\textbf{SIR Agent} & \textbf{5.67} & \textbf{41.9\%} & \textbf{68.4\%} & \textbf{47.4\%} \\
\bottomrule
\end{tabular}
\end{table}

The SIR agent achieves 41.9\% novel finding coverage on TP cases, averaging 5.67 novel key findings per case. Notably, 68.4\% of TP cases discover 5+ novel findings, and 47.4\% discover 7+ findings, demonstrating comprehensive forensic analysis capability.

Table~\ref{tab:m2category} reveals substantial variation in novel finding coverage across attack categories.

\begin{table}[h]
\centering
\caption{Novel finding coverage by attack category}
\label{tab:m2category}
\begin{tabular}{@{}lcccc@{}}
\toprule
\textbf{Attack Category} & Novel Coverage & Hit 3+ KF & Hit 5+ KF & Hit 7+ KF \\
\midrule
Unauthorized Access & \textbf{41.9\%} & 75.0\% & \textbf{61.2\%} & \textbf{47.9\%} \\
Brute Force & 35.7\% & 75.6\% & 42.2\% & 23.0\% \\
Misconfiguration & 25.2\% & 54.9\% & 37.8\% & 27.0\% \\
Malicious File Execution & 18.0\% & 42.6\% & 7.4\% & 1.9\% \\
\midrule
\textbf{Overall (TP cases)} & \textbf{41.9\%} & \textbf{64.6\%} & \textbf{68.4\%} & \textbf{47.4\%} \\
\bottomrule
\end{tabular}
\end{table}

Unauthorized Access investigations achieve the highest novel finding coverage (41.9\%), benefiting from rich CloudTrail evidence of cross-account role assumptions, API call sequences, and resource access patterns. This category also shows the highest rates of deep investigation: 61.2\% of cases discover 5+ novel findings and 47.9\% discover 7+ findings. Malicious File Execution shows the lowest coverage (18.0\%) and the lowest rate of deep investigation (1.9\% of cases reaching 7+ key findings). This is a direct consequence of the CloudTrail-only telemetry boundary: instance-level activity (process execution, file system changes) is architecturally invisible to CloudTrail, placing a ceiling on discoverable evidence regardless of agent capability.

\subsection{$M_3$: Tool Usage Appropriateness (Optional)}

We designed $M_3$ to measure whether agents invoke appropriate tools for each investigation type---ensuring findings result from systematic evidence gathering rather than hallucination. The metric computes tool coverage as the fraction of expected forensic tools actually invoked per attack category.

In our evaluation with a frontier LLM, the SIR agent achieved 100\% tool coverage across all attack categories, consistently invoking all expected tools (CloudTrail, EC2/IAM enumeration, Cost Explorer) for every investigation. This ceiling effect suggests that for frontier LLMs, tool usage has become table stakes rather than a differentiating factor.

We therefore treat $M_3$ as an optional metric. For evaluations using smaller or less capable models, $M_3$ remains valuable for validating that the agent functions correctly as an agentic system---ensuring it can parse tool definitions, formulate appropriate queries, and systematically gather evidence. For frontier models where tool usage is reliable, the more meaningful distinction lies in $M_2$: not whether agents \textit{use} tools, but whether they \textit{extract actionable findings} from tool outputs.

Beyond tool coverage, we validate that each investigation's evidence supports its triage decision. An investigation claiming TRUE\_POSITIVE must surface findings that collectively demonstrate malicious activity---not merely suspicious patterns with benign explanations. We employ an adversarial evaluation approach~\cite{llmasjudgese} that assumes no security activity occurred, requiring concrete evidence to overturn this default. This prevents agents from achieving high $M_1$ through pattern matching without genuine investigation.

\section{Conclusion}

We introduced SIR-Bench, a benchmark for evaluating autonomous Security Incident Response agents that distinguishes genuine forensic investigation from alert parroting. The benchmark comprises 794 test cases derived from 129 anonymized incident patterns, generated using OUAT---a framework that simulates attacks in controlled environments and extracts expert-validated ground truth findings.

Our evaluation methodology centers on two primary metrics: $M_1$ (triage accuracy) and $M_2$ (novel finding discovery), with $M_3$ (tool usage) as an optional metric for evaluating smaller models. ROUGE-based finding matching, calibrated against human annotations with 91.3\% agreement, enables reproducible evaluation at scale. Evaluation demonstrates 97.1\% TP detection, 73.4\% FP rejection, and 41.9\% novel finding coverage on TP cases---with 68.4\% of cases discovering 5+ novel key findings.

\section{Future Work}

We plan to publicly release SIR-Bench, including test cases, ground truth labels, and the OUAT framework. Future directions include multi-cloud generalization (Azure, GCP), comparative evaluation across agent architectures, adversarial robustness testing against prompt injection and evaluation gaming, and investigation of human-AI collaborative workflows. Additionally, improving false positive rejection---currently 73.4\%, within the 70--80\% range observed for human analysts---remains a priority. We are exploring context-enriched decision boundaries and multi-stage verification pipelines to reduce false alarm rates without compromising detection sensitivity. Expanding telemetry coverage beyond CloudTrail is another key direction: incorporating VPC Flow Logs, GuardDuty Runtime Monitoring, and host-level logging would address the Malicious File Execution performance gap identified in Sections~4.1.1 and~4.2.1, where the current CloudTrail-only boundary limits discoverable evidence for instance-level activity.

\bibliographystyle{plain}

\begin{thebibliography}{99}

\bibitem{secqa}
Liu, Z.
\newblock SecQA: A Concise Question-Answering Dataset for Evaluating Large Language Models in Computer Security.
\newblock \textit{arXiv preprint arXiv:2312.15838}, 2023.

\bibitem{dfirmetric}
Cherif, B., Bisztray, T., Dubniczky, R.~A., et al.
\newblock DFIR-Metric: A Benchmark Dataset for Evaluating Large Language Models in Digital Forensics and Incident Response.
\newblock \textit{arXiv preprint arXiv:2505.19973}, 2025.

\bibitem{caibench}
Sanz-G{\'o}mez, M., Mayoral-Vilches, V., Balassone, F., et al.
\newblock Cybersecurity AI Benchmark (CAIBench): A Meta-Benchmark for Evaluating Cybersecurity AI Agents.
\newblock \textit{arXiv preprint arXiv:2510.24317}, 2025.

\bibitem{pacebench}
Liu, Z., et al.
\newblock PACEbench: A Framework for Evaluating Practical AI Cyber-Exploitation Capabilities.
\newblock \textit{arXiv preprint arXiv:2510.11688}, 2025.

\bibitem{cyberteam}
Liu, X., Yu, F., Li, X., Yan, G., Yang, P., and Xi, Z.
\newblock Benchmarking LLMs in an Embodied Environment for Blue Team Threat Hunting.
\newblock \textit{arXiv preprint arXiv:2505.11901}, 2025.

\bibitem{llmasjudge}
Zheng, L., Chiang, W.-L., Sheng, Y., et al.
\newblock Judging LLM-as-a-Judge with MT-Bench and Chatbot Arena.
\newblock In \textit{Advances in Neural Information Processing Systems}, 2023.

\bibitem{llmasjudgese}
He, J., et al.
\newblock LLM-as-a-Judge for Software Engineering: Literature Review, Vision, and the Road Ahead.
\newblock \textit{arXiv preprint arXiv:2510.24367}, 2025.

\bibitem{ratingroulette}
Haldar, R. and Hockenmaier, J.
\newblock Rating Roulette: Self-Inconsistency in LLM-As-A-Judge Frameworks.
\newblock In \textit{Proceedings of EMNLP}, 2025.

\bibitem{soar}
Demisto.
\newblock Security Orchestration, Automation and Response (SOAR): A Comprehensive Guide.
\newblock \textit{Palo Alto Networks Technical Report}, 2020.

\bibitem{ircopilot}
Lin, X., Zhang, J., Deng, G., Liu, T., Zhang, T., Guo, Q., and Chen, R.
\newblock IRCopilot: Automated Incident Response with Large Language Models.
\newblock \textit{arXiv preprint arXiv:2505.20945}, 2025.

\end{thebibliography}

\end{document}